\documentclass[aps,prb,twocolumn,showpacs,floatfix]{revtex4}
\usepackage{graphicx}
\usepackage{amsmath}
\begin{document}

\title{The cluster glass state in the two-dimensional extended t-J model}
\author{Chung-Pin Chou, Noboru Fukushima, and Ting Kuo Lee}
\affiliation{Institute of Physics, Academia Sinica, NanKang, Taipei
11529, Taiwan}
\begin{abstract}
The recent observation of an electronic cluster glass state composed
of random domains with unidirectional modulation of charge density
and/or spin density on $Bi_{2}Sr_{2}CaCu_{2}O_{8+\delta}$
reinvigorates the debate of existence of competing interactions and
their importance in high temperature superconductivity. By using a
variational approach, here we show that the presence of the cluster
glass state is actually an inherent nature of the model based on the
antiferromagnetic interaction ($J$) only, {\it i.e.} the well known
$t-J$ model. There is no need yet to introduce a competing
interaction to understand the existence of the cluster glass state.
The long-range pairing correlation is not much influenced by the
disorder in the glass state which also has nodes and linear density
of states. In the antinodal region, the spectral weight is almost
completely suppressed. The modulation also produces subgap
structures inside the "coherent" peaks of the local density of
states.
\end{abstract}

%\keywords{}
\maketitle

%%%%%%%%%%%%%%%%%%%%%%%%%%%%%%%%%%%%%%%%%%%%%%%%%%%%%%%%%%%%%%%%%%%%
%                        Introduction
%%%%%%%%%%%%%%%%%%%%%%%%%%%%%%%%%%%%%%%%%%%%%%%%%%%%%%%%%%%%%%%%%%%%
\section{Introduction}
Since the discovery of high temperature superconductors (HTS) two
decades ago, many anomalous properties have been reported. One of
the most interesting properties is the possible existence of the
stripe state consisting of one dimensional charge-density modulation
coupled with spin ordering
\cite{ZaanenPRB89,PoilblancPRB89,KivelsonRMP03}. The first direct
experimental evidence \cite{TranquadaNat95} of this stripe state as
the ground state is for the non-superconducting
$La_{1.48}Nd_{0.4}Sr_{0.12}CuO_{4}$ with the hole density about
$1/8$ per unit cell. Since then more evidences of the presence of
these have been reported in other cuprate samples
\cite{MookPRL02,FujitaPRL02}. However it seems that the stripe is
much more prominent in the LaSrCuO (LSCO) family near $1/8$ doping
\cite{LakeNature02,LucarelliPRL03} and in particular for
$La_{1.875}Ba_{0.125}CuO_{4}$ ($LBCO-1/8$) \cite{FujitaPRL02} where
the charge density or spin density modulation could be considered as
an long-ranged order parameter of a phase with broken symmetries.
Very recently the high resolution scanning tunneling microscopy
(STM) has observed unidirectional domains with periodic density of
states modulation in two families of $Ca_{2-x}Na_{x}CuO_{2}Cl_{2}$
and $Bi_{2}Sr_{2}CaCu_{2}O_{8+\delta}$ (BSCCO)
\cite{KohsakaSci07,HowardPRB2003}. These bond-centered electronic
patterns with a width of four lattice constants form the so called
electronic cluster glass state with short-ranged modulations.
Although the modulation is weak, it is still surprising to have the
rather high superconducting (SC) transition temperature in this
glass state. STM spectra also provided the new puzzling result that
there are two different types of spectral gaps in those systems. One
is a larger gap with a broader distribution and it seems to be
related to the pseudogap. Inside this gap, there is a smaller gap or
the so called sub-gap kink with clear $d$-wave like spectra
\cite{BoyerNP07,AlldredgeCM08}. These results and the most recent
conflicting results reported by angle-resolved-photo-emission
spectra (ARPES) \cite{WSLeeNature07,KanigelPRL07} raise the question
whether there are two energy scales related with two different
underlying mechanisms separately responsible for the larger
pseudogap and the lower-temperature SC transition. But one feature
that most experimental results agree is the presence of the linear
density of state (DOS) near the nodal point. It is interesting to
note that this has also been reported in the non-superconducting
stripe state of $LBCO-1/8$ \cite{VallaSci06}.

The observations of these strongly-modulated inhomogeneous states
like $LBCO-1/8$ with almost zero SC transition temperature ($T_c$)
or weakly inhomogeneous cluster glass states in BSCCO with quite
high $T_c$ have fueled the idea about the presence of competing
interactions and underdoped HTS is near the boundary of two distinct
phases with different order parameters. The fluctuations
\cite{KivelsonRMP03} of the order parameters in these adjacent
phases could be the mechanism of high temperature superconductivity.
Thus it seems like a $d\acute{e}j\grave{a}$ $vu$ that after two
decades we are still faced with a daunting question about the
appropriate fundamental interactions to model and to understand the
HTS. Many people are still not convinced by all those successes
\cite{AndersonScience87,AndersonJPCM04,PALeeRMP06} claimed by
studying the strong coupling Hubbard model or its equivalent $t-J$
model. In this model the nearest-neighbor antiferromagnetic (AF)
spin coupling $J$ is the only relevant interaction besides the usual
kinetic energy of electrons (represented by the $t$ term). There are
no other competing interactions and certainly no phase boundaries
between overdoped and underdoped regimes with different order
parameters to be worried.

In this paper we will show that interactions represented by $t$ and
$J$ are actually competing with each other. This competition,
greatly enhanced by the strong correlation between electrons, has
many spatially heterogeneous states almost exactly the same energy
as the uniform ground state. The ground state could easily tolerate
local spatial modulation of charge density, spin density and even
pairing amplitude without much an effect on its SC order parameter.
The presence of these very low-energy cluster glass states with a
random pattern of short-ranged modulation is an inherent nature of
the $t-J$ model. The selection of a particular local electronic
pattern as observed in scanning tunneling spectroscopy (STS) for HTS
is most likely determined by the effects of impurities, defects and
electron-lattice interaction, etc. Random distribution of impurities
and defects will not produce long-ranged modulation in the sense of
charge-density-wave order or spin-density-wave order unless these is
a very strong lattice distortion as demonstrated \cite{FujitaPRL02}
by the structural transition observed in $LBCO-1/8$. Hence our
result for the extended $t-J$ model shows that for weakly
inhomogeneous cuprates like BSCCO there is no need to introduce
other strong competing interactions with new broken symmetry phases
and long-ranged order parameters to produce the observed glass
states. The extended $t-J$ model is adequate to explain many of the
experimental observations. Depending on the particular type of
modulations, the local DOS could either have a node with linear DOS
or without a node. These local modulations also produce  the sub-gap
structures. In addition, these cluster glass states have almost
completely suppressed the quasi-particle spectral weight near the
antinodal region as observed by ARPES for BSCCO compounds
\cite{TanakaSci06}.

Before we start with the discussion about the cluster glass states,
we should point out that the competition between the kinetic energy
represented by $t$ and the magnetic energy represented by $J$ is not
a new idea. It is the main reason for the uniform-state phase
diagram obtained by the theory of the resonating-valence-bond (RVB)
state \cite{AndersonScience87, AndersonJPCM04,PALeeRMP06}. The $J$
term prefers the formation of spin pairing, the so called RVB
singlet, or the long-ranged antiferromagnetism. Because of the
strong correlation a spin can only hop by exchanging position with a
hole, hence the kinetic energy is proportional to the density of
holes. But the more hole the system has, the less number of spin
there is. Consequently the pairing is also reduced. Thus the pairing
amplitude reduces when hole doping increases. Magnetic energy is
reduced while kinetic-energy gain increases. It is shown below this
competition also happens in individual unit cell.

On the theoretical side the stripe state also has a long history. It
was first founded almost two decades ago in the mean-field treatment
of the Hubbard model \cite{ZaanenPRB89,PoilblancPRB89} although the
validity of the method in treating strong Hubbard interaction is
questionable. Then the extended $t-J$ models were studied by several
numerical methods such as the exact diagonalization method
\cite{HellbergPRL99}, the density matrix renormalization group
method \cite{WhitePRB98,TohyamaPRB1999} and the variational Monte
Carlo (VMC) method \cite{TohyamaPRB1999,KobayashiJLTP1999}. But the
results are inconsistent. There are indications that stripe is
unstable when the second neighbor hopping, $t'$, is included in the
$t-J$ model. However, a later VMC study of the $t-t'-J$ model
\cite{HimedaPRL02} indicates that the stripe has about $1\%$ lower
energy than the uniform RVB based $d$-wave SC state for most of the
negative values of $t'$ except when $-t'/t$ is less than $0.1$.
Similar results were also reported for the Hubbard model
\cite{MiyazakiJPCS02}. These results pose a new direct contradiction
with the experimental findings. Many experimental and theoretical
studies have found that $-t'/t$ is smallest for LSCO family
\cite{PavariniPRL01,ShihPRL04,Hashimoto2008}, thus the stripe state
should not be favorable. Yet as mentioned above, among all the
cuprates LSCO family has the most solid evidences for the existence
of stripe. There are also other problems with the proposed stripe
states, such as suppressed pairing correlation and absence of the
V-shape DOS at low energy, and we will discuss them below. All of
the previous works concentrated on studying periodic stripes with a
long-ranged order, it is unclear if the result will hold for the
cluster glass state. There are also different kinds of stripes to be
considered. It is possible to have only charge density modulation,
or only spin density modulation, or only pairing amplitude
modulation or with the linear combination of any two or all three of
them \cite{MillisPRB07,Baruch08012436}. Furthermore the relation
between these modulations could be correlated or anti-correlated.
All these issues are addressed below and their results are compared
with the experiments.

%%%%%%%%%%%%%%%%%%%%%%%%%%%%%%%%%%%%%%%%%%%%%%%%%%%%%%%%%%%%%%%%%%%%
%                          VMC Part
%%%%%%%%%%%%%%%%%%%%%%%%%%%%%%%%%%%%%%%%%%%%%%%%%%%%%%%%%%%%%%%%%%%%
\section{The stripe-like states by the variational Monte Carlo method}
We consider the extended $t-J$ Hamiltonian,
\begin{eqnarray}
H=-\sum_{i,j,\sigma}t_{ij}\left(\tilde{c}_{i\sigma}^{\dag}\tilde{c}_{j\sigma}+h.c.\right)+J\sum_{<i,j>}\mathbf{S}_{i}\cdot\mathbf{S}_{j}
\label{e:Equ1}
\end{eqnarray}
which has been used to describe many physical systems in the
high-temperature superconductors \cite{KYYangPRB06}. The hopping
amplitude $t_{ij}=t$, $t'$, and $t''$ for sites $i$ and $j$ being
the nearest-, the second-nearest, and the third-nearest-neighbors,
respectively. We restrict the electron creation operators
$\tilde{c}_{i\sigma}^{\dag}$ to the subspace with no-doubly-occupied
sites. $\mathbf{S}_{i}$ is the spin operator at site $i$ and $<i,j>$
means that the interaction occurs only for the nearest-neighboring
sites. In the following, we mainly focus on the case $t''=-t'/2$ and
$J/t=0.3$ at hole doping $1/8$.

We shall follow the work by Himeda {\it et al.} \cite{HimedaPRL02}
to construct the variational wave functions. In the mean-field
theory we assume a local AF order parameter, the staggered
magnetization $m_{i}$, and nearest neighbor pairing order parameter
$\Delta_{ij}$. Thus the effective mean-field Hamiltonian is reduced
to
\begin{equation}
H_{MF}=\sum_{i,j}\left(\begin{array}{cc}
c^{\dag}_{i\uparrow} & c_{i\downarrow} \\
\end{array}\right)\left(\begin{array}{cc}
H_{ij\uparrow} & D_{ij} \\
D^{\ast}_{ji} & -H_{ji\downarrow} \\
\end{array}\right)\left(\begin{array}{c}
c_{j\uparrow} \\
c^{\dag}_{j\downarrow} \\
\end{array}\right),
\label{e:Equ2}
\end{equation}
where the matrix elements
\begin{eqnarray}
H_{ij\sigma}=&-&\left(t_{v}\sum_{\beta=N}+t'_{v}\sum_{\beta=NN}+t''_{v}\sum_{\beta=NNN}\right)\delta_{j,i+\beta}\nonumber\\
&+&\left(\rho_{i}-\mu_{v}+\sigma(-1)^{x_{i}+y_{i}}m_{i}\right)\delta_{j,i},
\label{e:Equ3}
\end{eqnarray}
\begin{equation}
D_{ij}=\sum_{\beta=N}\Delta_{ij}\delta_{j,i+\beta}. \label{e:Equ4}
\end{equation}
Here $\beta=N$, $NN$, and $NNN$ correspond to the nearest-, the
next-nearest, and the third-nearest-neighbors, respectively, and
$\sigma=\uparrow$(1) or $\downarrow$(-1). The local charge density
is controlled by $\rho_{i}$ and $\mu_{v}$ is the variational
parameter for the chemical potential. For periodic stripes we assume
charge density $\rho_{i}$ and staggered magnetization $m_{i}$ are
anti-correlated, ${\it i.e.}$ there are more holes at sites with
minimum staggered magnetization. For simplicity, we assume these
spatially varying functions with simple forms:
\begin{equation}
\rho_{i}=\rho_{v}\cos[4\pi\delta\cdot(y_{i}-y_{0})], \label{e:Equ5}
\end{equation}
\begin{equation}
m_{i}=m_{v}\sin[2\pi\delta\cdot(y_{i}-y_{0})], \label{e:Equ6}
\end{equation}
where $\delta$ is the doping density and is $1/8$ in this paper. We
have also taken the lattice constant to be our unit. Here we assume
the stripe extends uniformly along the $x$ direction. $y_{0}=0$
$(1/2)$ corresponds to the site- (bond-) centered stripe. In this
paper, we will only focus on the bond-centered stripe since the
variational energy difference between the site- and bond-centered
stripes is very small (not shown) within the finite cluster
\cite{RaczkowskiPRB07,CapelloPRB08}.

Besides staggered magnetization and  charge density, the nearest
neighbor pairing $\Delta_{ij}$ can also have a spatial modulation.
There are several different stripes we can choose. If $\Delta_{ij}$
has the same period $1/\delta$ as the staggered magnetization but it
is $\pi$ phase shifted, this is the so called "antiphase" stripe
studied by a number of groups
\cite{HimedaPRL02,RaczkowskiPRB07,CapelloPRB08,HimedaJPCS02}. In
this stripe state, the bond-average $\Delta_{ij}$ is zero and there
is no net pairing. Hole density is maximum at the sites with maximum
pairing amplitude $|\Delta_{ij}|$ and minimum staggered
magnetization $|m_i|$.

Another more general choice is to have the spatial variation of
$\Delta_{ij}$ of the form,
\begin{eqnarray}
\Delta_{i,i+\hat{x}}&=&\Delta^{M}_{v}\cos[4\pi\delta\cdot(y_{i}-y_{0})]-\Delta^{C}_{v},\nonumber\\
\Delta_{i,i+\hat{y}}&=&-\Delta^{M}_{v}\cos[4\pi\delta\cdot(y_{i}-y_{0})+2\pi\delta]+\Delta^{C}_{v}.
\label{e:Equ7}
\end{eqnarray}
The bond-average $\Delta_{ij}$ is determined by the constant
$\Delta^{C}_{v}$. If both $\Delta^{M}_{v}$ and $\Delta^{C}_{v}$ are
positive, then the hole density is maximum at sites with smallest
pairing amplitude $|\Delta_{ij}|$ and smallest magnetization
$|m_i|$. This is similar to the phase diagram
\cite{ShihLTP05,ChouJMMM07} predicted by the uniform RVB and AF
states, when hole density is small both staggered magnetization and
pairing amplitude are larger. Thus we will denote this state as the
AF-RVB stripe. For AF-RVB stripe the period of $\Delta_{ij}$ is the
same as the charge-density modulation $\rho_{i}$. This period,
$1/2\delta$, is only half of the period for the antiphase or $\pi$
phase stripe. Besides the antiphase stripe and AF-RVB stripes, we
could also have the AF stripe without both $\Delta^{M}_{v}$ and
$\Delta^{C}_{v}$ or the charge-density stripe without any staggered
magnetization but with pairing amplitude modulation.

In general we have total seven variational parameters $\mu_{v}$,
$t'_{v}$, $t''_{v}$, $\rho_{v}$, $m_{v}$, $\Delta^{M}_{v}$, and
$\Delta^{C}_{v}$ with $t_{v}$  set to be 1. Once these parameters
are given, we diagonalize the mean-field Hamiltonian in equation
(\ref{e:Equ2}). By solving the Bogoliubov de Gennes (BdG) equations
\begin{equation}
\sum_{j}\left(\begin{array}{cc}
H_{ij\uparrow} & D_{ij} \\
D^{\ast}_{ji} & -H_{ji\downarrow} \\
\end{array}\right)\left(\begin{array}{c}
u^{n}_{j} \\
v^{n}_{j} \\
\end{array}\right)=E_{n}\left(\begin{array}{c}
u^{n}_{i} \\
v^{n}_{i} \\
\end{array}\right),
\label{e:Equ8}
\end{equation}
and then obtain $N$ positive eigenvalues $E_{n}$ ($n=1-N$) and $N$
negative eigenvalues $\bar{E}_{n}$ with corresponding eigenvectors
$(u^{n}_{i},v^{n}_{i})$ and $(\bar{u}^{n}_{i},\bar{v}^{n}_{i})$. The
eigenvectors are used to construct the mean-field wave function
\cite{HimedaPRL02} $|\psi\rangle$ by using Bogoliubov transformation
\begin{equation}
\left(\begin{array}{c}
\gamma_{n} \\
\bar{\gamma}_{n} \\
\end{array}\right)=\left(\begin{array}{cc}
u^{n}_{i} & v^{n}_{i} \\
\bar{u}^{n}_{i} & \bar{v}^{n}_{i} \\
\end{array}\right)\left(\begin{array}{c}
c_{i\uparrow} \\
c^{\dag}_{i\downarrow} \\
\end{array}\right).
\label{e:Equ9}
\end{equation}
The trial wave function $P|\psi\rangle$ with a Gutzwiller projector
$P$ can be constructed by creating all negative energy states and
annihilating all positive energy states on a vacuum of electrons
$|0\rangle$. Then we formulate the wave function in the Hilbert
space with the fixed number of electrons $N_{e}$,
\begin{eqnarray}
|\Phi\rangle=PP_{N_{e}}|\psi\rangle&=&PP_{N_{e}}\prod_{n}\gamma_{n}\bar{\gamma}^{\dag}_{n}|0\rangle\\\nonumber
&\propto&
P\left(\sum_{i,j}(\hat{U}^{-1}\hat{V})_{ij}c^{\dag}_{i\uparrow}c^{\dag}_{j\downarrow}\right)^{N_{e}/2}|0\rangle,
\label{e:Equ10}
\end{eqnarray}
where $\hat{U}_{ij}=u^{i}_{j}$ and $\hat{V}_{ij}=v^{i}_{j}$. We
optimize the variational energy
$E=\langle\Phi|H|\Phi\rangle/\langle\Phi|\Phi\rangle$ by using the
stochastic reconfiguration algorithm \cite{SorellaPRB01}.
Additionally, to reduce the boundary-condition effect in numerical
studies \cite{PoilblancPRB91}, we average the energies over the four
different boundary conditions which is periodic or antiperiodic in
either $x$ or $y$ direction.

\begin{figure}[top]
\rotatebox{0}{\includegraphics[height=3.0in,width=3.2in]{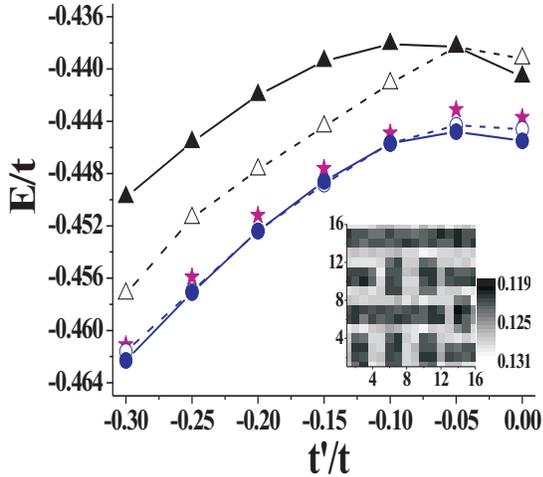}}
\caption{Variational energies of the bond-centered AF-RVB stripe
state and uniform $d$-wave RVB state as function of $t'/t$. The data
have been optimized for $1/8$ doping on the $16\times16$ lattice
system. The filled (empty) symbols represent the energies of uniform
$d$-wave RVB state (AF-RVB stripe state). The triangles (circles)
indicate the wave functions without (with) the hole-hole correlation
of equation (\ref{e:Equ11}). The asterisks correspond to the random
stripe states with the Jastrow factor of equation (\ref{e:Equ11}).
\textbf{Inset}: The hole density of the random stripe state
optimized in the case of $(t',t'',J)/t=(-0.3,0.15,0.3)$.}
\label{f:Fig1}
\end{figure}

\begin{figure}[top]
\rotatebox{0}{\includegraphics[height=5.5in,width=3in]{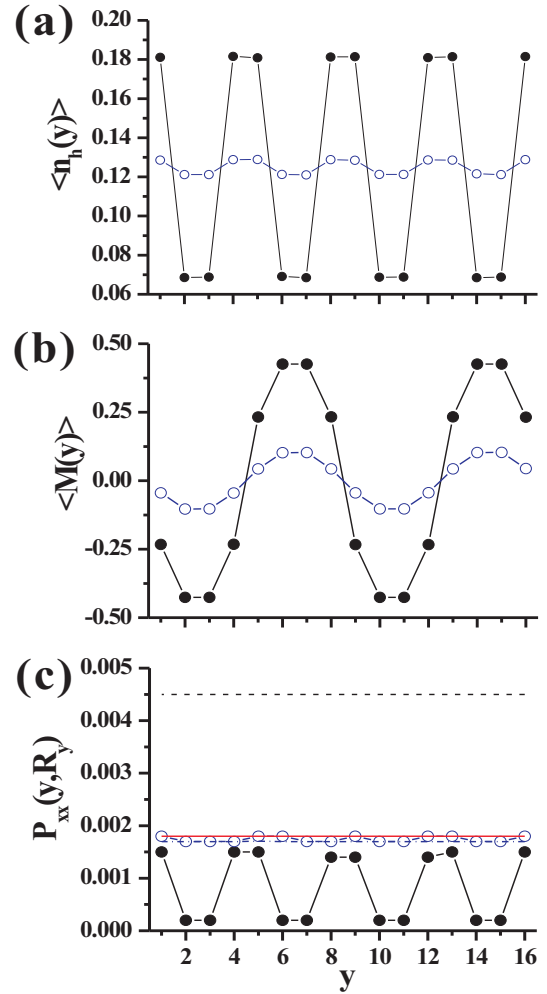}}
\caption{The profiles of (a) hole density, (b) staggered
magnetization and (c) pair-pair correlation function with $R_{y}=8$
for the optimized states at $1/8$ doping in the extended $t-J-$
model with $t'/t=-0.3$ and $J/t=0.3$. The filled (empty) circles
correspond to the AF-RVB stripe state without (with) hole-hole
repulsion. In (c), the dashed (dashed-dotted) line indicates the
uniform $d$-wave RVB state without (with) hole-hole repulsion. The
red solid line corresponds to the random stripe state with hole-hole
repulsive Jastrow correlation. All quantities are calculated on a
$16\times16$ lattice system with the periodic and antiperiodic
boundary conditions along the x and y directions, respectively.}
\label{f:Fig2}
\end{figure}

Fig.\ref{f:Fig1} shows the $t'/t$ dependence of the variational
energies for the hole density $\delta=1/8$. Firstly, we have shown
that the AF-RVB stripe state (the empty triangles) becomes more
stable than uniform $d$-wave RVB state (the filled triangles) as
decreasing $t'/t$ further from $-0.05$. It is worth noting that the
AF-RVB stripe state for $t'/t<-0.05$ has the vanishing pairing
parameters $\Delta^{C}_{v}$ and $\Delta^{M}_{v}$ but large $m_{v}$
and finite $\rho_{v}$. Therefore, we can consider this stripe state
without SC order as the antiphase AF stripe state
\cite{ZaanenPRB89}. The hole density and the staggered magnetization
are plotted as a function of positions for a typical AF stripe
(filled circles) in Fig.\ref{f:Fig2}(a) and (b), respectively. The
AF stripe pattern with a very large hole-density variation is
essentially a nano-scale phase separation with hole-rich and
hole-poor regions alternating. This is consistent with what has been
obtained by Himeda {\it et al.} \cite{HimedaPRL02} although they did
not include $t''$ which is set to be $-t'/2$ here.

In our previous calculations to study the possibility of phase
separation in the $t-J$ model \cite{ShihJPCS2001}, we found the
tendency to overestimate the strength of the pairing of holes. If we
reduce this strength by going beyond the simple trial wave functions
used in our discussion above we could push the phase separation
boundary to a much higher value of $J/t$. A simpler way to make this
adjustment is to introduce the hole-hole repulsion Jastrow
factor\cite{HellbergPRL91,ValentiPRL92,SorellaPRL02}:

\begin{equation}
P_{J}=\prod_{i<j}\left(1-(1-r^{\alpha}_{ij}\cdot
v^{\delta_{j,i+\beta}}_{\beta})\cdot n^{h}_{i}n^{h}_{j}\right)
\label{e:Equ11}
\end{equation}
with
\begin{equation}
r_{ij}=\sqrt{\sin^{2}\left(\frac{\pi}{L_{x}}(x_{i}-x_{j})\right)+\sin^{2}\left(\frac{\pi}{L_{y}}(y_{i}-y_{j})\right)},\nonumber
\end{equation}
where $n^{h}_{i}=1-\Sigma_{\sigma}c^{\dag}_{i\sigma}c_{i\sigma}$.
The three parameters $v_{\beta}$ of $\beta=N$, $NN$, and $NNN$ are
for short-ranged hole-hole repulsion (attraction) if these values
are less (greater) than 1. The factor $r^{\alpha}_{ij}$ is for
long-ranged correlations \cite{HellbergPRL91} and it is repulsive if
$\alpha$ is positive. $L_{x}$ and $L_{y}$ are the number of sites in
the $x$ and $y$ direction, respectively. We have found that the
variational energy of the periodic stripe state without including
the Jastrow factors is very sensitive to boundary conditions.
However the Jastrow factor is capable of reducing the dependence of
the boundary conditions.

As shown in Fig.\ref{f:Fig1}, when the hole-hole repulsion of
equation (\ref{e:Equ11}) is included in our VMC calculation, both
the uniform $d$-wave RVB state (filled circles) and the AF-RVB
stripe state (empty circles) have gained significant energies. In
particular, the uniform $d$-wave RVB state has gained about
$2\sim3\%$ in total energy. It indicates that for the extended $t-J$
model the holes are not bounded as tightly as estimated by the
simple RVB trial wave functions. It also implies that the nano-scale
phase separation of the AF stripe will not be preferable. Indeed,
the lowest energy stripe state does not favor the AF stripe but the
AF-RVB stripe with a much more reduced staggered magnetization as
shown in Fig.\ref{f:Fig2}(b) and a larger pairing parameter
$\Delta^{C}_{v}$. Now the hole density has a much smaller variation
as shown by the empty circles in Fig.\ref{f:Fig2}(a).

It is surprising to find out that for all $t'/t$, the AF-RVB stripe
state is almost degenerate in variational energy with the uniform
$d$-wave RVB state. This is quite remarkable as the two trial wave
functions are very different. As an illustration we show in
Fig.\ref{f:Fig2}(a) and (b) the variation of the hole density and
the staggered magnetization along the modulation direction,
respectively, for an AF-RVB stripe state (empty circles). It should
be noticed that not only the hole density is completely uniform for
$d$-RVB state but there is also no staggered magnetization at all.
Instead of periodic stripes, we have also examined the stripe state
with $4\times4$ patches in the $16\times16$ lattice system. For each
patch, we choose a direction of the stripe, $x$ or $y$, randomly. We
consider this state as random stripe state. For simplicity, we still
use the same equations (\ref{e:Equ5}-\ref{e:Equ7},\ref{e:Equ11}).
The hole-density modulation of a random stripe state has been shown
in the inset of Fig.\ref{f:Fig1}. As shown in Fig.\ref{f:Fig1}, the
random stripe states with the Jastrow factors of equation
(\ref{e:Equ11}) also have the optimized energies almost identical to
uniform $d$-wave RVB and AF-RVB stripe states even though we have
not optimized parameters on very bond or site independently. These
optimized random stripe states have finite $m_{v}$ and
$\Delta^{C}_{v}$ but smaller $\Delta^{M}_{v}$ (less than one third
of $\Delta^{C}_{v}$).

We believe this energy degeneracy is caused mainly by two reasons.
The first reason is that the terms in the $t-J$ Hamiltonian are all
local within nearest neighbors or next nearest neighbors. The second
reason is that the hopping terms and spin interaction not only are
of the same order of magnitude but they are competing against each
other. We have found the energy competition between the kinetic
energy and spin interaction is very robust among different states
(not shown). Their competition is enhanced by the no-doubly
occupancy constraint as the presence of holes will suppress the spin
interaction to zero. Thus it is possible to have locally different
spin-hole configurations with different emphasis on the kinetic
energy or the spin energy. Some of these patterns have lower kinetic
energy but higher magnetic energy than the uniform $d$-RVB state and
some with opposite energetics.

There are at least two important implications of this energy
degeneracy. The first one is that the inhomogeneous states are quite
robust in the extended $t-J$ models without the need for introducing
any other large interactions. States with different local
arrangement of spin and holes may have very similar energies. The
second implication is that any small additional interaction could
break the degeneracy. If we consider the "realistic" situation of
cuprates with large numbers of impurities, disorders and
electron-lattice interactions, the inhomogeneous states could be
much more numerous and complex than we have expected. Materials made
with different processing conditions could also produce different
inhomogeneous states. Since all these are presumably secondary
interactions smaller than the dominant $t$ and $J$, the modulations
of charge density, staggered magnetization and pairing are expected
to be small. If one of the interactions, like electron-lattice
interaction, becomes quite strong as observed in $LBCO-1/8$
\cite{VallaSci06}, then the modulation will also become larger and
longer ranged.

We have also investigated the pair-pair correlation function for the
optimized states with/without hole-hole repulsive Jastrow factors in
the case of $(t',t'',J)/t=(-0.3,0.15,0.3)$. The singlet pair-pair
correlation function along the modulated direction (y direction) is
defined as
\begin{equation}
P_{xx}(y,R_{y})=\frac{1}{L_{x}}\sum_{x}\frac{|\langle\Phi|\Delta^{\dag}_{x}(r)\Delta_{x}(r+R_{y})|\Phi\rangle|}{\langle\Phi|\Phi\rangle},
\label{e:Equ12}
\end{equation}
where $r=(x,y)$ and
$\Delta^{\dag}_{x}(r)=c^{\dag}_{r\uparrow}c^{\dag}_{r+\hat{x}\downarrow}-c^{\dag}_{r\downarrow}c^{\dag}_{r+\hat{x}\uparrow}$
creates a singlet pair of electrons among the nearest neighbors
along $x$ direction for each site $r$. Here, we focus on the long
range correlation, and thus set $R_{y}=8$ to be the largest distance
on $16\times16$ lattice system. In Fig.\ref{f:Fig2}(c), it is shown
that the hole-hole repulsion suppresses the long-range pair-pair
correlation about three times in the uniform $d$-wave RVB state. We
have also found that without hole-hole repulsive correlation, the
long-range pair-pair correlation is much more reduced in the AF-RVB
stripe state than uniform $d$-wave RVB state, because the AF-RVB
stripe state has almost vanishing $\Delta^{C}_{v}$. Due to large
$m_{v}$, this AF-RVB stripe state shows much larger amplitude of the
$P_{xx}$ modulation. However, after considering the hole-hole
repulsion, the AF-RVB stripe and uniform $d$-wave RVB states have
almost the similar magnitude of pair-pair correlation as shown by
the empty circles and dashed-dotted line in Fig.\ref{f:Fig2}(c),
respectively. For the random stripe state shown in the inset of
Fig.\ref{f:Fig1} the average value $P_{xx}$ for the whole system is
shown as the red solid line in Fig.\ref{f:Fig2}(c). It is
essentially the same as the value of uniform $d$-wave RVB state and
the periodic AF-RVB stripe state. Although the staggered
magnetization for the random stripe state has larger variation than
the periodic stripe state shown in Fig.\ref{f:Fig2}(b), the pairing
correlation is unchanged. These results indicate that the long-range
pair-pair correlation is mostly determined by the value of
$\Delta^{C}_{v}$ and is rather insensitive to the modulation of the
hole density and staggered magnetization. Thus we also expect to
have a robust $d$-wave node observed by recent experiments
\cite{KohsakaSci07,AlldredgeCM08}. This will be shown below.

%%%%%%%%%%%%%%%%%%%%%%%%%%%%%%%%%%%%%%%%%%%%%%%%%%%%%%%%%%%%%%%%%%%%
%                          GA Part
%%%%%%%%%%%%%%%%%%%%%%%%%%%%%%%%%%%%%%%%%%%%%%%%%%%%%%%%%%%%%%%%%%%%
\section{Density of states by the Gutzwiller approximation}
According to the VMC calculation for the extended $t-J$ model, it is
likely that there are a number of inhomogeneous states close in
energy to the uniform ground state. Then, some sort of small
perturbation may choose a particular stripe state as the ground
state. Assuming such a situation, here we regard a stripe state as
the ground state, and consider the projected quasi-particle
excitation spectra. However, calculation of the excited states by
the VMC method is computationally very expensive, and it takes too
much time to investigate wide parameter range to obtain general
properties of the stripe states. Furthermore, one can take only a
limited system size and it is difficult to obtain dense spectra.

Therefore, as a first step, we use a Gutzwiller mean-field
approximation for this purpose. The minimization of the total energy
yields a BdG equation \cite{Fukushima08012280}. Usually the
parameters in the BdG equations are solved self-consistently to find
an optimal solution. However, since we already have the assumed
inhomogeneous ground state here, we shall use parameter sets
obtained from VMC results and diagonalize the BdG Hamiltonian only
once, instead of solving self-consistently. Furthermore, for
convenience, we slightly modify the Gutzwiller projection by
attaching fugacity factors. Namely, we assume that
$P_{\lambda}|\psi\rangle$ is the ground state and that
$P_{\lambda}\gamma_n^\dagger|\psi\rangle$ are the excited states,
where $P_{\lambda}\equiv P
\prod_{i\sigma}\lambda_{i\sigma}^{n_{i\sigma}}$, and
$\lambda_{i\sigma}$ is a fugacity factor to impose the local
electron density conservation for each spin, namely,
$\frac{\langle\psi|P_{\lambda}\hat{n}_{i\sigma
}P_{\lambda}|\psi\rangle} {\langle\psi|P_{\lambda}^2|\psi\rangle}=
\frac{\langle\psi|\hat{n}_{i\sigma }|\psi\rangle} {\langle
\psi|\psi\rangle}$ for any $i$ and $\sigma$. The quasi-particle
operators $\gamma_n^\dagger$ are obtained by solving the BdG
Hamiltonian. In addition, here we do not take into account the
Jastrow factor as it only affects the hole-hole correlation slightly
but not the local DOS studied below. Then, under the assumption of
the non-self-consistency, the BdG Hamiltonian is represented by
equation (\ref{e:Equ2}). With this formulation,
$P_{\lambda}\gamma_n^\dagger|\psi\rangle$ of different $n$ are
approximately orthogonal to each other \cite{Fukushima08012280}, and
thus we expect that it is suitable to use $P_{\lambda}$ instead of
$P$ for our purpose here. Since the result presented below is
qualitatively not very sensitive to small change of the parameters,
we expect that such a modification of the projection should not
affect the results qualitatively.

\begin{figure}[top]
\includegraphics[width=4cm]{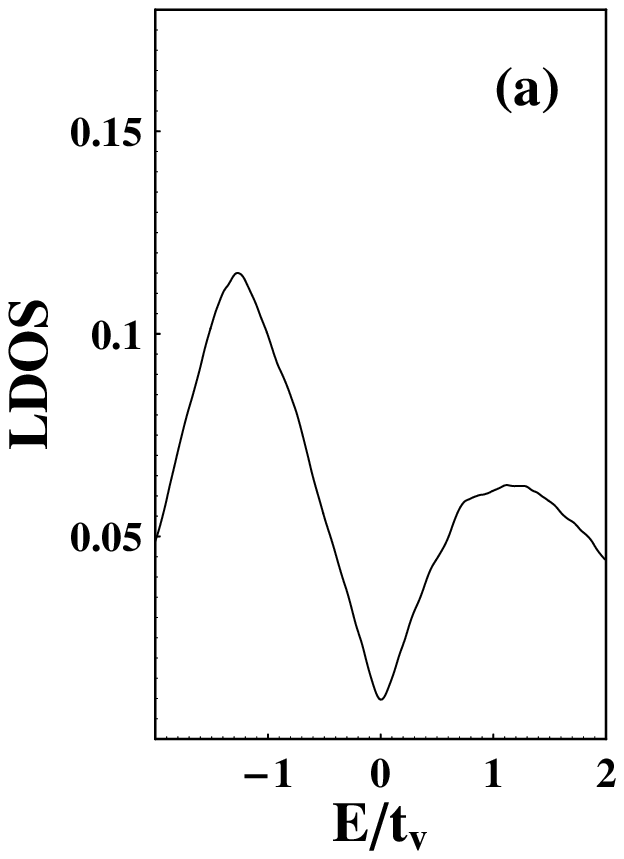}
\includegraphics[width=4cm]{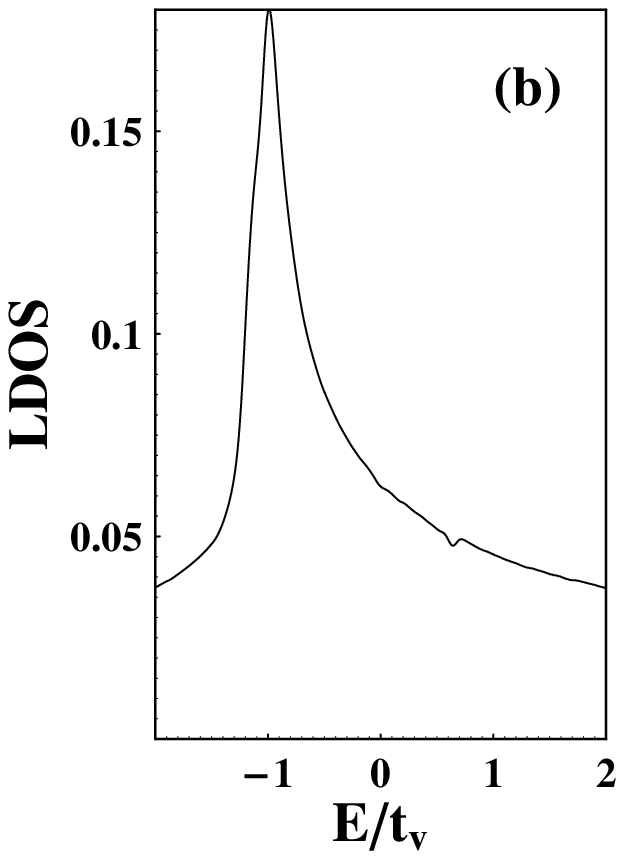}
\caption{Spatially averaged local DOS for the random stripe states
calculated by the non-self-consistent BdG equation. (a) For
$\Delta_v^C\neq0$, we use parameters optimized by the VMC,
$t'_v=-0.35$, $t''_v=0.16$, $m=0.15$, $\rho=0.03$, $\Delta_v^C=0.28$
, $\Delta_v^M=0.02$, but $\mu= - 0.875t_v$ is adjusted to realize
$1/8$ filling, in units of $t_v$. (b) The same parameters except for
$\Delta_v^C = 0$. \label{fig:randomGutzLDOS}}
\end{figure}

Then, by taking the most dominant terms, the local DOS is
represented by
\begin{gather}
N_{\uparrow}(R,\omega) = g^t_{R\uparrow}\sum_n |u_R^n|^2
\delta(\omega-E_n),
\\
N_{\downarrow}(R,\omega) = g^t_{R\downarrow}\sum_n |v_R^n|^2
\delta(\omega+E_n) ,
\end{gather}
where index $n$ runs for both positive and negative eigenvalues.
Note that only the position dependent constant $g^t_{R\sigma}\equiv
{( 1 - n_{R\uparrow} - n_{R\downarrow} )}/ {( 1 - n_{R\sigma} )}$ is
multiplied in front of the local DOS by the standard BdG formalism.
Since the result of the site-centered stripe is very similar to that
of the bond-centered stripe, we show only the latter. Here we shall
only discuss our results for the random stripe state. For the random
stripe state considered above, each randomly oriented domain is
assumed to have the same parameters so that the VMC calculation is
possible. Ideally we should have optimized these variational
parameters on every bond or every site. Then we expect to have a
much broader distribution of these parameters. To simulate this
effect,  we simply replace each $\Delta_{ij}$ by
$(1+\xi_{ij})\Delta_{ij}$, where $\xi_{ij}$ is a random variable
which has the Gaussian distribution around 0 with the standard
deviation of 1. We use a supercell of size 32$\times$32 sites, and
the same configuration is repeated as 20$\times$20 supercells to
obtain the local DOS. The Fourier transform with respect to the
supercell index is similar to a system of small clusters with many
twisted boundary conditions.

\begin{figure}[top]
\includegraphics[width=8cm]{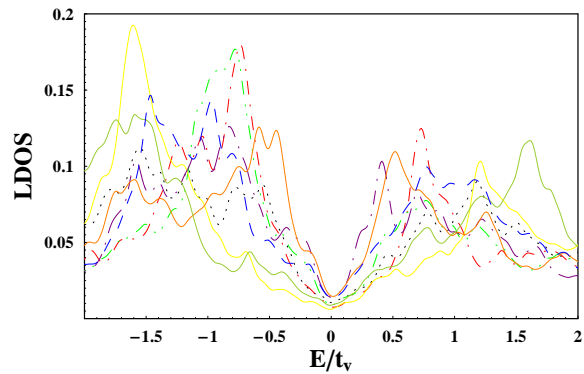}
\caption{Local DOS at eight different positions for the random
stripe states with $\Delta_v^C\neq0$ calculated by the
non-self-consistent BdG equation. \label{fig:LDOSposdep}}
\end{figure}

As shown in Fig.\ref{fig:randomGutzLDOS}(a), the spatially averaged
$\sum_\sigma N_\sigma(R,\omega)$ for the AF-RVB stripe
($\Delta_v^C\neq 0$) has a V-shape at low energy. On the other hand
for an antiphase stripe which has $\Delta_v^C=0$ there is no V-shape
as shown in Fig.\ref{fig:randomGutzLDOS}(b). The energy level is
broadened with a width of $0.05t_v$. In
Fig.\ref{fig:randomGutzLDOS}(b), there is a very small dip at $E=0$,
it would be bigger if $\Delta_v^M$ increases. Our results are
consistent with the very recent report by Baruch and Orgad
\cite{Baruch08012436}. In general, there is no V-shape DOS for the
antiphase stripe. For the same system as
Fig.\ref{fig:randomGutzLDOS}(a), in Fig.\ref{fig:LDOSposdep} we plot
the position dependence of local DOS at randomly chosen 8 sites. The
low energy spectra seem less influenced by the disorder than high
energy. This result shows that the node and the low energy V-shape
DOS are robust against this kind of inhomogeneity. This is possible
because nodal $k$-points do not have many states to mix with and
also the suppression of impurity scattering \cite{note1}. The
sub-gap structures \cite{BoyerNP07, AlldredgeCM08} also seem to be
quite apparent. If we switch off the random variables $\xi_{ij}$,
the gap variations from site to site are small. These gap variations
grow larger as distributions of $\xi_{ij}$ become wider.

The key to understand the absence of V-shape in the Local DOS for
the antiphase stripe is the Fourier transform of the modulated
$\Delta_{ij}$ term written as,
\begin{multline}
\Delta_v^M \sum_k \cos k_x (e^{-i\theta} c_{k+q,\uparrow}^\dagger
c_{-k,\downarrow}^\dagger + e^{i\theta} c_{k-q,\uparrow}^\dagger
c_{-k,\downarrow}^\dagger + \text{h.c.})
\\
-\frac{\Delta_v^M}{2}\sum_k \bigg[
 \left( e^{i k_y }  +  e^{-i (k_y+q_y) } \right)
e^{i(\frac{q_y}{2}-\theta)}
 c_{k+q,\uparrow}^\dagger c_{-k,\downarrow}^\dagger
\\
+
 \left( e^{i k_y }  +  e^{-i (k_y-q_y) } \right)
e^{-i(\frac{q_y}{2}-\theta)} c_{k-q,\uparrow}^\dagger
c_{-k,\downarrow}^\dagger + \text{h.c.}\bigg] ,
\end{multline}
where $\theta=0$ for site-centered stripes and $\theta=q_y/2$ for
bond-centered stripes; $q=(0,\pi/4)$ for the antiphase stripe, and
$q=(0,\pi/2)$ for the AF-RVB stripe. What is important here is that
it contains {\it only} pairing with {\it nonzero} center-of-mass
momentum as the FFLO state \cite{QWangPRL06}. In the case of
zero-momentum pairing as the conventional BCS theory, the spin-up
electron band couples with the spin-down hole band (the upside-down
down electron band). These bands intersect at the Fermi level, and a
gap opens if $\Delta_k\neq 0$. In the case of finite-$q$ pairing,
however, the spin-up electron band couples with $\pm q$ shifted
spin-down hole bands, and thus the band intersections occur not at
the Fermi level. Therefore, a gap does not open at the Fermi level.
The constant $\Delta^{C}_{v}$ term which forms the usual Cooper pair
is necessary for having the node and the V-shape DOS.

To compare with ARPES experiments, $A(k,\omega)$ is also calculated.
Since $A(k,\omega)$ is regarded as the local DOS in the $k$-space,
Let us take the Fourier transform of {\it renormalized} $u_R^n$,
$v_R^n$, namely,
\begin{equation}
( \tilde{u}_k^n , \tilde{v}_k^n ) \equiv \frac{1}{\sqrt{N_{\rm
site}}} \sum_R e^{- i k R} \left( g^t_{R\uparrow} u_R^n,
g^t_{R\downarrow} v_R^n \right).
\end{equation}
Then, $A_{\sigma}(k,\omega)$ is written as
\begin{gather}
A_{\uparrow}(k,\omega) = \sum_n |\tilde{u}_k^n|^2
\delta(\omega-E_n),
\\
A_{\downarrow}(k,\omega) = \sum_n |\tilde{v}_k^n|^2
\delta(\omega+E_n).
\end{gather}

\begin{figure}[top]
\includegraphics[width=3.5cm]{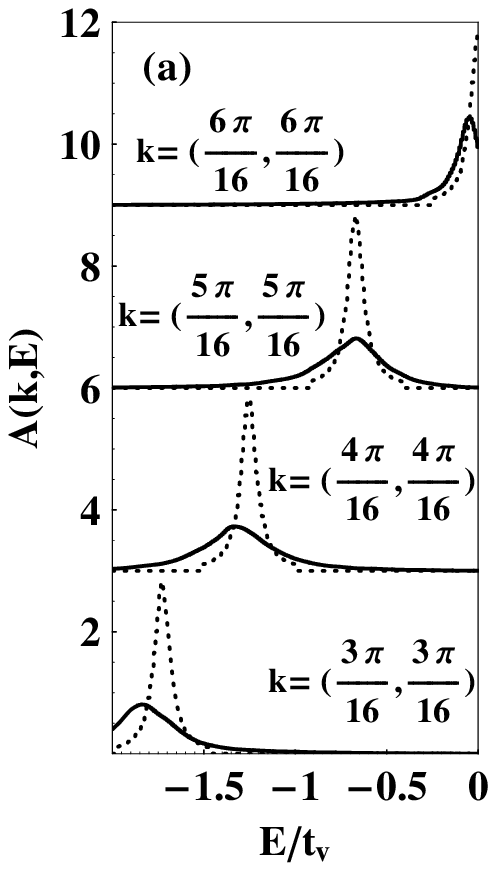}
\includegraphics[width=3.5cm]{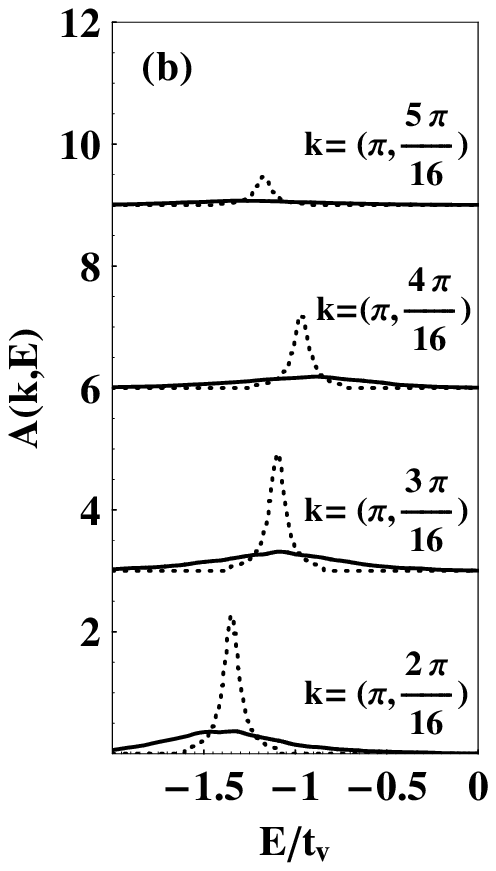}
\caption{$\sum_\sigma A_\sigma(k,E)$ for the random stripe state
with $\Delta_{C}\neq0$. Dotted lines are for uniform $d$-wave RVB
state ($\rho_v=0, m_v=0,\Delta_v^M=0$). (a) Nodal and (b) near
antinodal region. \label{fig:AFRVBAkwr}}
\end{figure}

Fig.\ref{fig:AFRVBAkwr} shows $\sum_\sigma A_\sigma(k,\omega)$ of
the random stripe state with $\Delta_v^{C}\neq0$, where each of the
$\delta$-function spectra are replaced with a Lorentzian
distribution with $\delta E = 0.05t_v$. In comparison with the
uniform $d$-wave RVB state, the spectral weight around the antinodal
region is much strongly reduced than near the node.

%%%%%%%%%%%%%%%%%%%%%%%%%%%%%%%%%%%%%%%%%%%%%%%%%%%%%%%%%%%%%%%%%%%%
%                         Conclusion
%%%%%%%%%%%%%%%%%%%%%%%%%%%%%%%%%%%%%%%%%%%%%%%%%%%%%%%%%%%%%%%%%%%%
\section{Conclusions}
In summary, we have used a variational approach to examine the
possibility of having inhomogeneous ground states within the
extended $t-J$ model with $1/8$ doping. We considered states with
spatial modulation of charge density, staggered magnetization and
pairing amplitude. Besides the antiphase or inphase stripes
considered by many groups we have proposed a new AF-RVB stripe. In
this stripe state, we assume there is a constant pairing amplitude
besides the various modulation. In addition to considering states
with periodic stripes we also consider random stripe states to
simulate the cluster glass state observed by experiments
\cite{KohsakaSci07}. By improving the trial wave functions with the
introduction of hole-hole repulsive correlation, we have greatly
improved the variational energies by several percents for both
uniform RVB $d$-wave SC state and states with periodic AF-RVB
stripes for realistic values of $t'/t$. Most surprisingly the random
stripe state essentially also has the same energy as the uniform
state in spite of our oversimplified assumption that all the stripe
domain has the same patterns of modulation instead of each site or
bond with different values. This random stripe state also has about
the same long-range pair-pair correlation as the uniform or periodic
stripe state even though there are significant staggered
magnetization and charge variation from site to site. Then we also
examined the local DOS and the spectral weight of the random stripe
state by using Gutzwiller approximation. We found the V-shape DOS
and the node are still present at every site. The local DOS measured
at different positions shows a broad variation of the gaps and also
it has sub-gap structures seen in experiments
\cite{BoyerNP07,AlldredgeCM08}. The spectral weight at the antinodal
direction is negligibly small but finite around the node. All these
results are quite consistent with experiments reported for cluster
glass state in BSCCO \cite{KohsakaSci07}.

Our result also resolves an inconsistency with experiments derived
from previous theoretical calculations without including the
hole-hole repulsion in the trial wave functions. Stripe is neither
stabilized nor destabilized by the long range hopping. In fact, due
to the competition between the kinetic energy gain and magnetic
interaction, it is very natural to have the spatial modulation, in
periodic or random configuration, of charge density, magnetization
and even pairing amplitude. The constraint of disallowing doubly
occupation of electrons at each lattice site has significantly
enhanced the competition. Many local arrangements of spin and hole
configurations could give almost identical total energy as the
uniform solution.

Recently, Capello {\it et al.}
\cite{CapelloPRB08} have also found that the energy of the periodic
RVB  stripe state is very close to that of the uniform RVB state by using
a variational calculation. They have considered several possibilities
for the stability of the  RVB stripe state, such as lattice
distortion, t'-effect, and long-range Coulomb repulsion with the conclusion that
uniform RVB state is still the lowest energy state. This is very consistent with our conclusion although
we have included the antiferromagnetic order in the stripe state. The issue about whether AF is present  in
the stripe will be discussed in the future. They have not considered the
 cluster glass state with random AF-RVB stripe domains which is also a good candidate for the
ground state.

The presence of inhomogeneous or cluster glass states is apparently
a very natural consequence of the $t-J$ model. There is no need for
introducing additional interactions to generate such states. In fact
we showed that the RVB state with a finite constant pairing is quite
compatible with the local variations of charge density,
magnetization and even pairing amplitude. As long as this modulation
is not overly strong, the superconductivity still survives as the
node and V-shape DOS are still present. In a realistic material,
other interactions such as impurity, disorder, and electron-lattice
interactions, etc., no doubt will help to determine the most
suitable local configuration of spins and holes but they will not
produce a globally ordered state unless there is a very strong and
dominant interaction like the electron-lattice interaction seen in
$La_{2-x}Ba_{x}CuO_{4}$ at $1/8$ doping. The verification of this is
left for future work.

\section{Acknowledgments}
This work is supported by the National Science Council in Taiwan
with Grant no.95-2112-M-001-061-MY3. The calculations are performed
in the IBM Cluster 1350, FormosaII Cluster and IBM P595 in the
National Center for High-performance Computing in Taiwan, and the PC
Farm and Euler system of Academia Sinica Computing Center in Taiwan.


\begin{thebibliography}{99}
\bibitem{note1}Due to the projection operator, the impurity scattering matrix element is strongly renormalized with a factor proportional to the hole denstiy.
\end{thebibliography}
\end{document}